\documentclass[a4paper,twoside]{article}
\usepackage[authoryear]{natbib}
\bibpunct{(}{)}{;}{a}{}{,}

\usepackage{graphicx}
\usepackage{times}
\usepackage{url}
\date{} %Please leave the date blank
\title{\large\bf\flushleft Automated Editing of Radio Interferometer Data with Pieflag}
\author{\parbox{\textwidth}{\flushleft
\vspace{-0.5cm}
{\it Enno Middelberg}\\
\vspace{0.4cm}
{\small Australia Telescope National Facility, PO Box 76, Epping NSW 1710, Australia}\\
{\small Email: enno.middelberg@csiro.au}}}
\begin{document}
\twocolumn[
\begin{changemargin}{.8cm}{.5cm}
\begin{minipage}{.9\textwidth}
\vspace{-1cm}
\maketitle
%
%
%%%%%%%%%%%%%     ABSTRACT    %%%%%%%%%%%%%
%Abstract of no more than 200 words here.
\small{\bf Abstract:}

Editing of radio interferometer data, a process commonly known as
``flagging'', can be laborious and time-consuming. One quickly tends
to flag more data than actually required, sacrificing sensitivity and
image fidelity in the process. I describe a program, Pieflag, which can
analyse radio interferometer data to filter out measurements which are
likely to be affected by interference. Pieflag uses two algorithms to
allow for data sets which are either dominated by receiver noise or by
source structure. Together, the algorithms detect essentially all
affected data whilst the amount of data which is not affected by
interference but falsely marked as such is kept to a minimum. The
sections marked by Pieflag are very similar to what would be deemed
affected by the observer in a visual inspection of the data. Pieflag
displays its results concisely and allows the user to add and remove
flags interactively. It is written in Python, is easy to install and
use, and has a variety of options to adjust its algorithms to a
particular observing situation. I describe how Pieflag works and
illustrate its effect using data from typical observations.

%%%%%%%%%%%%%     KEYWORDS    %%%%%%%%%%%%%
\medskip{\bf Keywords:} 
% Please write all keywords in lower case. PASA uses the
% standard list of subject headings adopted by The Astrophysical Journal
% and available from http://www.journals.uchicago.edu/ApJ/keywords_text.html.
% Keywords are separated by em-dashes, i.e. ---
methods: data analysis

%%%%%%%%DO NOT EDIT%%%%%%%%%%%%
\medskip
\medskip
\end{minipage}
\end{changemargin}
]
\small
%%%%%%%%EDIT FROM HERE%%%%%%%%%%%%

\section{Introduction}
%Please see the PASA Style Guide for help with correct layout for your manuscript.
%Examples of tables and figures are given below.

Data editing, to rid data sets from measurements which are obviously
wrong (for whatever reason) or affected by external factors, is a
widely accepted practice in all branches of science. Radio
astronomical measurements, in particular at frequencies below 5\,GHz,
can be severely affected by emission from man-made transmitters, which
is then known as radio frequency interference (RFI). Only a few
attempts have been made as yet to automate the editing of radio
astronomical observations, an example includes the ``WSRT flagger''
used at the Westerbork
telescope\footnote{\url{http://www.astron.nl/~renting/flagger.html}}. However,
it is recognized that powerful RFI mitigation techniques need to be
developed for future instruments such as the Square Kilometre Array
(SKA) \citep{Ellingson2004}.

In radio interferometry, data editing is commonly known as
``flagging''. RFI-affected data are not discarded but an entry (a
``flag'') is made in a computer file, to indicate that the data are
not to be used in further processing. When large amounts of data have
to be handled, data flagging can be very laborious and dull. This can
lead to situations in which one either flags too many or too few data,
sacrificing good data in the first case, or accepting a degradation of
the quality of the observations in the latter.

The automation of flagging requires the creation of a measure that
indicates what is ``good'' data, and then the observed data needs to
be compared to this measure. Both of these requirements have to work
in a variety of circumstances, and assumptions which may work well in
one case can be invalid in another.

In Pieflag\footnote{The original name of the program, ``Pyflag'',
turned out to be already in use, requiring a name modification of some
sort to avoid confusion.}, two algorithms are implemented to detect
RFI-affected data made under a variety of circumstances. Two steps of
postprocessing extrapolate the flags which were created by Pieflag to
points close in time. This yields a safety margin around RFI-affected
data and picks up those data which are likely to be affected as well,
but have not been found by the finding algorithms.

Pieflag has been designed to read data processed with Miriad
\citep{Sault1995}, which is the standard data reduction package for
observations made with the Australia Telescope Compact Array
(ATCA). However, it will work on $(u,v)$ data from other
interferomters as well, when these are converted into the Miriad
format. I note that the algorithms presented here are not specific to
interferometric observations. Time series of single-dish radio
telescope data could be analysed in the same fashion.

I will first describe the details of the algorithms used to detect RFI
and to generate flags (Section~\ref{sec:works}), then the display of
the results and how the flags are applied to data sets
(Section~\ref{sec:display} and \ref{sec:apply}), followed by an
assessment of the limitations (Section~\ref{sec:limits}), and a
description of usage (Section~\ref{sec:usage}) and some examples
(Section~\ref{sec:examples}).

\section{How Pieflag works}
\label{sec:works}

Pieflag reads in data from Miriad data sets and tries to identify
RFI-affected sections of data by comparing the visibility amplitudes
of each frequency channel to a channel in the spectrum which is known
to contain little or no bad data (the reference
channel). Consequently, the data must be bandpass-calibrated before
Pieflag is used. The hypothesis used here is that data are affected
mostly by man-made RFI, which typically is limited to very narrow
frequency bands (see for example the spectra measured at the
ATCA\footnote{\url{http://www.narrabri.atnf.csiro.au/observing/rfi/}}). This
type of RFI thus can be detected when data from different channels are
compared. The amplitudes of atmospheric noise and the astronomical
signal usually vary between baselines and coordinates at which the
telescopes are directed (a ``pointing''), hence the comparison must be
done separately for each baseline and pointing.

The minimum number of parameters Pieflag needs to work is the name of a
directory with a Miriad data set and the number of the reference
channel. Pieflag works in four steps to generate flags, the first two
of which are used to determine which data are deemed to be
RFI-affected, and the last two of which transfer the flags to adjacent
data which are likely to be affected as well. These four steps are
described in the following sections.

\subsection{Step 1: Amplitude-based flagging}

In the first step, Pieflag computes the median visibility amplitude of
the reference channel, $x_{\rm b,p}$, for each baseline $b$ and
pointing $p$, and the median of the difference to this median, $y_{\rm
b,p}$. The use of the median and the median of the difference to the
median as a measure for the scatter protects the algorithm from
outliers in the reference channel. Both the mean and standard
deviation can easily be contaminated by single data points with very
high amplitudes, and are therefore not suitable. Pieflag then
calculates the difference of each data point in each channel to
$x_{\rm b,p}$. If the difference exceeds $ny_{\rm b,p}$, where $n$ is
typically around 7, a point is considered suspect and is assigned a
``badness'' value of 1. If the difference exceeds $2ny_{\rm b,p}$, it
is deemed certainly bad and is assigned a ``badness'' value of
2. Amplitude-based flagging is most efficient when the data are
dominated by receiver noise or in the presence of very weak
astronomical signals. In the presence of strong signals from
structured sources, $y_{\rm b,p}$ tends to be too high for the
algorithm to detect bad data which deviate only little from the source
signal (Figure~\ref{fig:n7462.bsl3-4.ch13}). The badness values are
converted to flags not before postprocessing
(Section~\ref{sec:post1}).

If the amplitudes of the visibility amplitudes in the reference
channel have a normal distribution, then 68.2\,\% of the data fall
within the range of $\pm 1\,\sigma$ of the mean. The median of the
difference to the median, however, chooses those 50\,\% of the data
that fall within the range of $x_{\rm b,p}\pm y_{\rm b,p}$, with
$y=0.674\sigma$ (this was computed numerically). Hence, in the case of
normally distributed amplitudes and $n=7$, $ny_{\rm b,p}$ is
equivalent to $n\times0.674\,\sigma=4.72\,\sigma$.

\subsection{Step 2: Rms-based flagging}

In this step, Pieflag calculates the median of the standard deviation,
or rms, in short (typically 2\,min to 3\,min) sections of reference
channel data for each baseline and pointing. The median rms value,
$z_{\rm b,p}$, is used as a measure of a typical rms to which the
other channels are compared. If the rms in a section exceeds $mz_{\rm
b,p}$, with $m$ typically 3, then the entire section is flagged. The
concept of badness is not used by this algorithm, because groups of
points are considered, and not single points. This algorithm can find
bad data even in the presence of a strong source signal, and therefore
complements the amplitude-based flagging. However, terrestrial and
solar interference can increase amplitude levels without much effect
on the rms on short timescales, which then is unnoticed by this
algorithm (but likely to be detected by the first).

\subsection{Step 3: Postprocessing to find small clusters of bad points}
\label{sec:post1}

If both amplitude- and rms-based flagging have been applied, Pieflag's
internal tables contain information about the badness of some points
as derived by amplitude-based flagging, and information about which
parts of the data should be flagged as derived from rms-based
flagging. To turn the badness values into flags, the running sum of
the badness values in a 1\,min window is calculated in each channel
and on each baseline. If the sum exceeds 1, the window is
flagged. This procedure will tolerate single, moderate outliers (one
value with badness 1 within one minute), but will flag the entire
window if two moderate outliers, or one point with badness 2, occur
within a minute. It also flags a small margin around the bad
data. This algorithm is applied irrespective of pointings, because it
is assumed that data are affected irrespective of changes of
pointings, at least on such short timescales.

\subsection{Step 4: Postprocessing to find larger clusters of bad points}

When developing the flagging algorithms, it was found that often a few
affected data points were not found during times with strong
interference. For example, some data points in
Figure~\ref{fig:bsl2-5.ch1} at 11:50\,h have amplitude and rms levels
which are not suspicious, given that most of the data in that channel
have similar amplitudes. However, considering the high levels of
interference immediately surrounding these data, one would certainly
flag them as well, if the editing was done manually.

One can argue that if any data has not been detected by the procedures
described above, they should not be flagged. Also, considering that
should these data be affected at all they would have only minuscule
effects on the final result, it seems pointless to flag them. On the
other hand, it is known in experimental sciences that bad data can be
worse than no data, and that discarding possibly good data as a safety
measure is acceptable (to some degree). I also point out that the
amount of data {\it additionally} flagged in this and the previous
step is of the order of a few percent at most, hence the sensitivity
loss is negligible. It therefore is a matter of personal taste and
level of caution whether or not to apply this step.

The flags are gridded into bins with a width of 30\,s and convolved
with a boxcar function with a width of typically 20\,min. The
amplitude of the convolution is a function of the fraction of data
which have been flagged in any period of 20\,min, and the maximum
possible value is predictable. If the fraction exceeds a threshold
(typically 0.15 of the maximum), the entire window is flagged. An
additional benefit of this procedure is that a larger safety margin is
flagged around extended periods of bad data. This step also is carried
out irrespective of pointings.\\

Except for step 3, each of the steps above is adjustable to one's
needs. Adjustable parameters are: the multiplication factors $n$ and
$m$ in the search for bad data, the width of the sections in which the
rms is calculated in rms-based flagging in step 2, and the width of
the boxcar function and the threshold above which the entire window is
flagged in step 4. Furthermore, steps 1, 2, and 4 can be skipped.

\section{Display of the results}
\label{sec:display}

\begin{figure}[htpb!]
\begin{center}
\includegraphics[width=\linewidth]{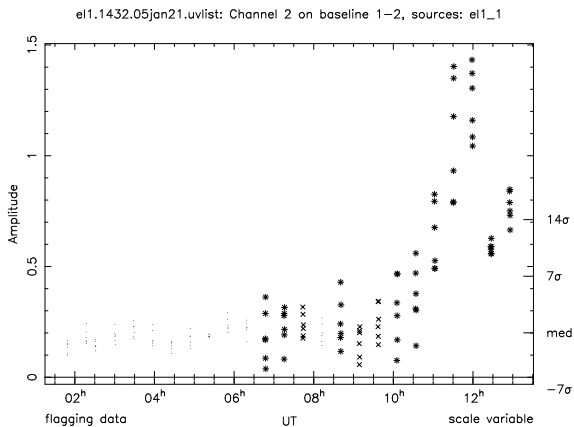}
\caption{An example of how Pieflag displays its results. The labels are
very small in print, but are large enough to read on a computer screen
for which they are intended. Amplitudes of one pointing, measured in
one frequency channel on one baseline are displayed. The plot
illustrates the two flagging algorithms. All data observed after
11:00\,h have amplitudes higher than $x_{\rm b,p}+7y$ or $x_{\rm
b,p}+14y$, indicated on the right of the plot. These points have been
flagged by amplitude-based flagging, followed by
postprocessing. Several scans between 6:00\,h and 11:00\,h have
amplitudes which are below the threshold of amplitude-based flagging,
but have been detected by rms-based flagging. Other scans between
6:00\,h and 11:00\,h have not been detected, but were flagged in
postprocessing, because data from adjacent pointings were detected and
flagged.}
\label{fig:screen}
\end{center}
\end{figure}

Pieflag displays its results to the user for iterative adjustment of
the parameters (Figure~\ref{fig:screen}). The visibility amplitudes of
one baseline and one channel are displayed, either from all pointings
at once, or from one pointing only. The latter is more instructive
when attempting to understand why data have been flagged. In the
single source mode, $x_{\rm b,p}$, $x_{\rm b,p}\pm ny$, and $x_{\rm
b,p}\pm 2ny$ are indicated on the right of the plot. Although the
plotting stage allows interactive flagging and unflagging using mouse
and keyboard controls, its primary intend is to allow one to inspect
the results. Consequently, once Pieflag is tuned to fit one's data, it
can be run without plotting.

%Good data are shown
%as dots, points which have been deemed affected by either amplitude-
%or rms-based flagging with plus symbols, and those data which have
%been finally flagged as crosses.

\section{Application of flags to the data}
\label{sec:apply}

Pieflag does not modify the flagging tables of Miriad data sets
directly. Instead, it writes a shell script to disk which has to be
executed to apply the flags. The script is only a sequence of Miriad
``uvflag'' commands with the appropriate selection of baselines,
channels, and times. The advantages of having a script are that the
flags can be re-applied later, without running Pieflag again, and that
a record is kept of how the flagging was done, in the form of comments
at the beginning of the script.

\section{Limitations}
\label{sec:limits}

Although Pieflag can deal with a variety of data sets, it should not be
used thoughtlessly.\\ 

If the amplitude on a baseline varies rapidly due to source structure,
then even the rms-based flagging cannot detect bad points. Either the
amplitude variations within each section will be high, making $z_{\rm
b,p}$ exceedingly large, or $z_{\rm b,p}$ is not representative
because the timescale of the amplitude variations changes largely
throughout the experiment.

%Furthermore, if the point-to-point amplitude variations due to source
%structure are large compared to the integration time, then one has to
%decrease the width of the sections in which the rms is
%calculated. This may result in an insufficient number of points to
%calculate a meaningful rms, and can result in false detections of good
%data, or in the non-detection of bad data.

The amplitude-based flagging assumes that the contribution of
astronomical sources to the signal is the same for all channels. This
assumption is not valid if the signal is dominated by sources with
large spectral indices. However, the rms-based flagging will be
unaffected by this.

However, both of these effects can be avoided if a suitable source
model, derived from the data in channels which are essentially free of
RFI, is subtracted from the data before Pieflag is used.\\

Pieflag should be used with caution in spectral line observations, when
the observed lines are strong. The channels containing line emission
would be flagged if compared to a line-free reference channel. In
spectral line detection experiments, however, Pieflag can be used
because the data are dominated by receiver noise. It should be noted
that spectral line observations are frequently much less affected by
terrestrial RFI than continuum observations, because the bandwidths
are much smaller and may be within protected parts of the
electromagnetic spectrum.  A further downside of using Pieflag on
spectral line data sets can be that the computing time increases
linearly with the number of channels. As a guideline, the computing
time for a 12\,h observation with all six ATCA antennae and a
correlator cycle time of 10\,s (the default mode) was found to be
6\,s per channel, using an otherwise idle 2.8\,GHz PC running Linux.

\section{Usage}
\label{sec:usage}

Details of where to get Pieflag, how to install and use it are
described on a
webpage\footnote{\url{http://www.atnf.csiro.au/people/Enno.Middelberg/pieflag}}. Pieflag
is expected to become part of the Miriad distribution soon.

\section{Examples}
\label{sec:examples}

In this section I describe a few examples of the use of
Pieflag. Detailed descriptions have been added to the image captions.

\begin{figure}[htpb!]
\begin{center}
\includegraphics[width=\linewidth]{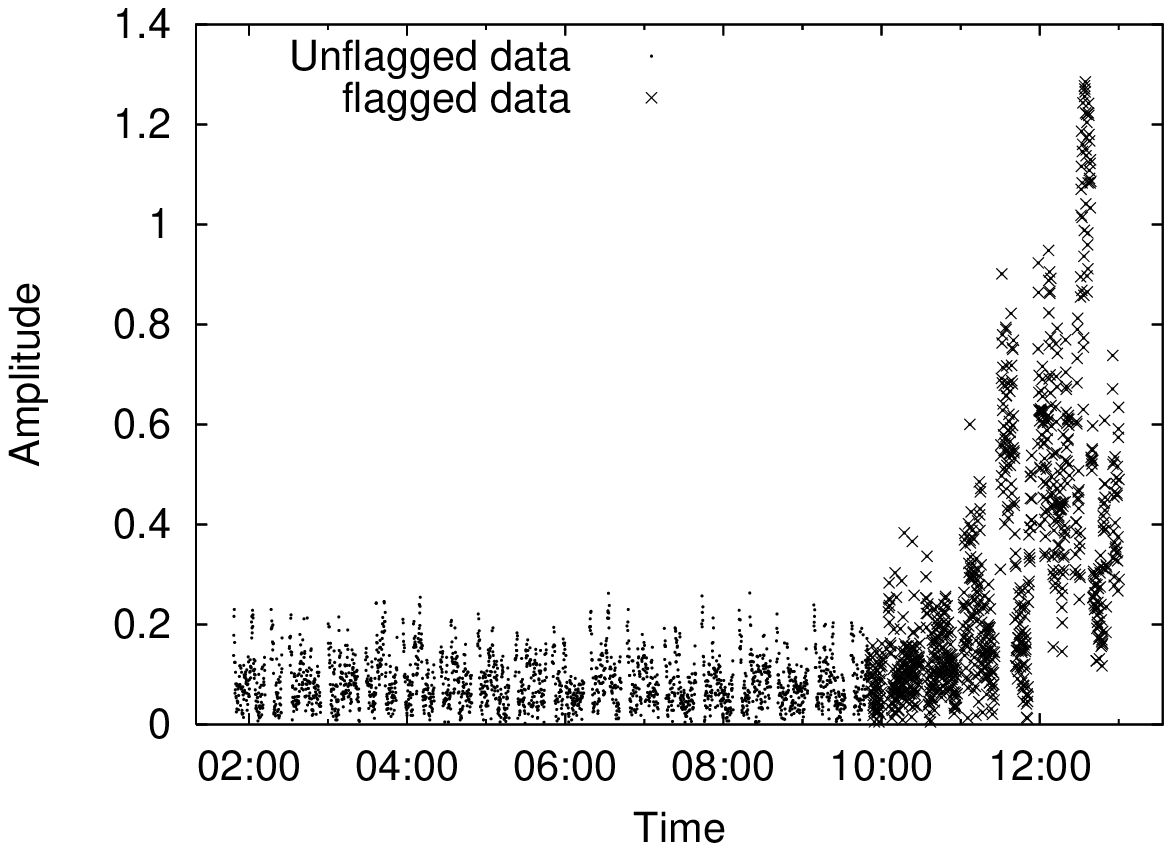}
\includegraphics[width=\linewidth]{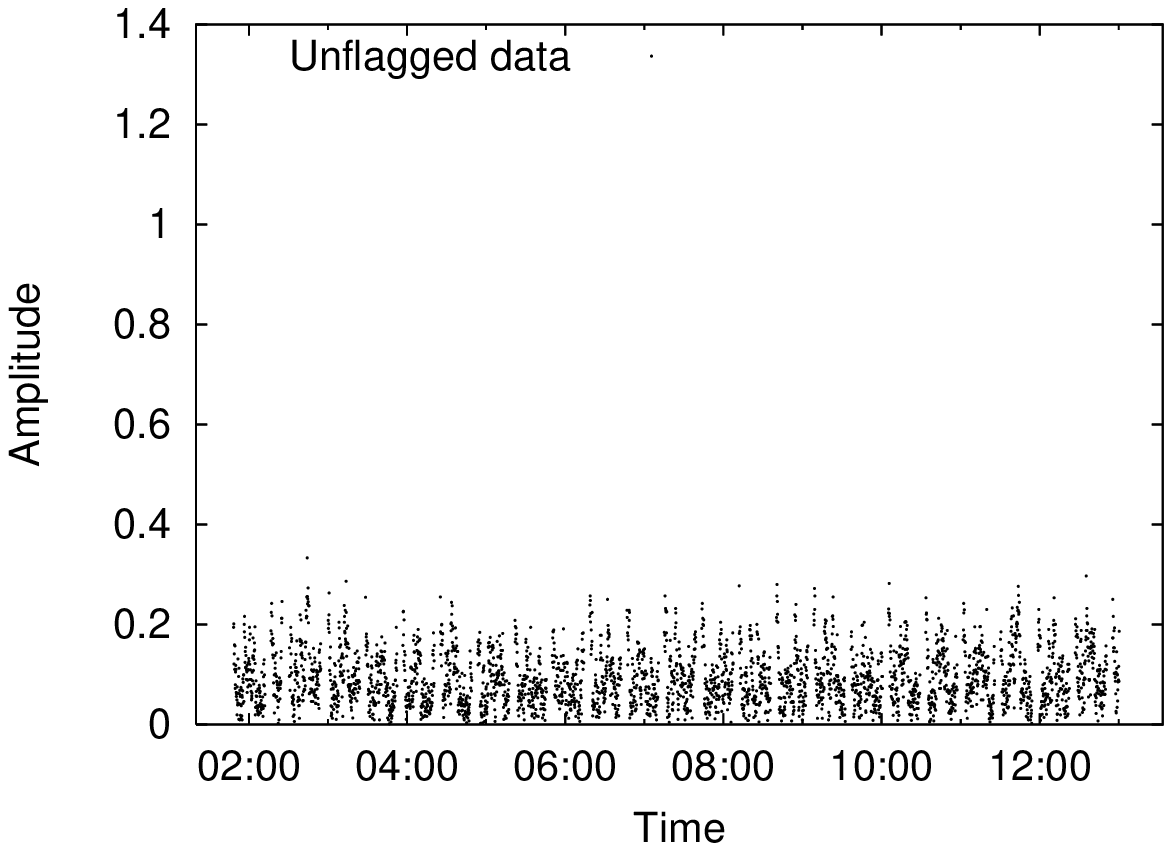}
\caption{{\it Top:} Visibility amplitudes of one frequency channel with
RFI on one baseline during an observation at 1.4\,GHz. Dots indicate
unflagged data, crosses indicate data which have been flagged by
Pieflag. 20 pointings were observed for one minute each, only one of
these has a flux density which is significantly higher than the
receiver noise, of around 0.2. After 10:00\,h, the data in this
channel are dominated by terrestrial RFI, whereas the reference
channel is RFI-free. Note the few data at 11:50\,h with amplitudes
below 0.2, which have been flagged by the second postprocessing step
because they are surrounded by bad data. {\it Bottom:} The visibility
amplitudes of the reference channel used to detect the bad data in
other channels. There are no obvious outliers or noisy sections.}
\label{fig:bsl2-5.ch1}
\end{center}
\end{figure}

Figure~\ref{fig:bsl2-5.ch1} shows the effect of Pieflag on data
observed with the ATCA at L-band outside the spectral regions reserved
for radio astronomy, between 1.38\,GHz and 1.48\,GHz. This band is
particularly contaminated by RFI at the beginning and end of
observations, when the antennae are pointed towards, or away from, the
town of Narrabri, at 20\,km distance from the telescope.

\begin{figure}[htpb!]
\begin{center}
\includegraphics[width=\linewidth]{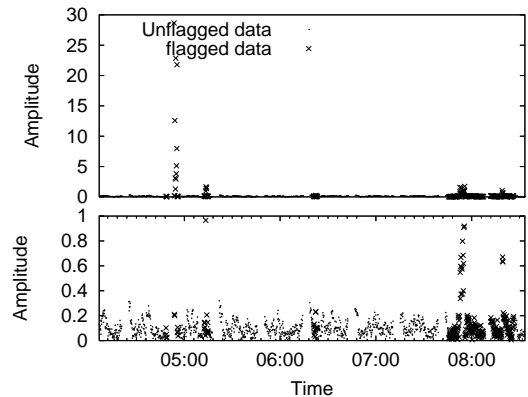}
\caption{Visibility amplitudes on the same baseline as in
Figure~\ref{fig:bsl2-5.ch1}, but showing another frequency channel
with a different kind of RFI, and only 4.5\,h of data. As the outliers
have large amplitudes, the amplitude range of 0 to 1 has been
magnified in the lower panel for clarity. This plot illustrates the
effect of the two postprocessing algorithms. Where only few outliers
have been detected, such as between 4:30\,h and 5:30\,h, only small
margins around the bad data were flagged during
postprocessing. However, between 7:45\,h and 8:30\,h, the fraction of
bad data within a 30\,min window exceeded the threshold of 0.15, and
extended margins were flagged. The gaps in the data are due to
calibrator observations.}
\label{fig:bsl2-5.ch13}
\end{center}
\end{figure}

Figure~\ref{fig:bsl2-5.ch13} shows data from the same observing run,
but from a different frequency channel. The RFI in this channel has
much higher amplitudes, but is confined to very short periods of
time. The figure illustrates the effect of the second step of
postprocessing.

\begin{figure}[htpb!]
\begin{center}
\includegraphics[width=\linewidth]{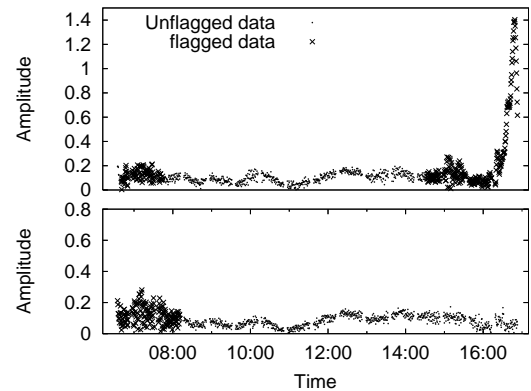}
\caption{{\it Top:} Visibility amplitudes of one channel on a baseline
observing a source with significant structure. The amplitudes
are therefore not noise-dominated but source-dominated, and the
amplitude-based flagging failed, except for the data taken after
16:30\,h, when the amplitude exceeded $x_{\rm b,p}+7y_{\rm
b,p}=0.28$. Two periods of RFI have been flagged by rms-based
flagging, one between 6:30\,h and 7:40\,h, and one between 15:00\,h
and 15:30\,h. Margins have been flagged around both of these periods
in postprocessing. {\it Bottom:} The reference channel on the same
baseline. The interference at the beginning of the experiment is also
present, but the usage of the median instead of the mean when
computing comparison values tolerates small amounts of bad data in the
reference channel. Data courtesy of M. Dahlem.}
\label{fig:n7462.bsl3-4.ch13}
\end{center}
\end{figure}

Figure~\ref{fig:n7462.bsl3-4.ch13} shows data from an experiment in
which the visibility amplitudes are dominated by the astronomical
source, which has significant structure. Flagging based on amplitude
did not work, since the cutoff was chosen such that much of the RFI at
the beginning of the experiment was not detected. However, rms-based
flagging did find the affected data. This figure also illustrates that
a small amount of RFI in the reference channel has no noticeable
effect on the flagging process.

\section{Conclusions}

Pieflag can efficiently search interferometry data for man-made
RFI. The implemented algorithms can deal with a variety of
observational circumstances, are robust and find virtually all
RFI-affected data which would be found by visual inspection. Usage of
Pieflag makes the flagging procedure much faster, relieves observers
from an annoying part of the data calibration, and can easily be set
up to run in data calibration scripts without any human interaction.

\begin{figure}[htpb!]
\begin{center}
\includegraphics[width=\linewidth]{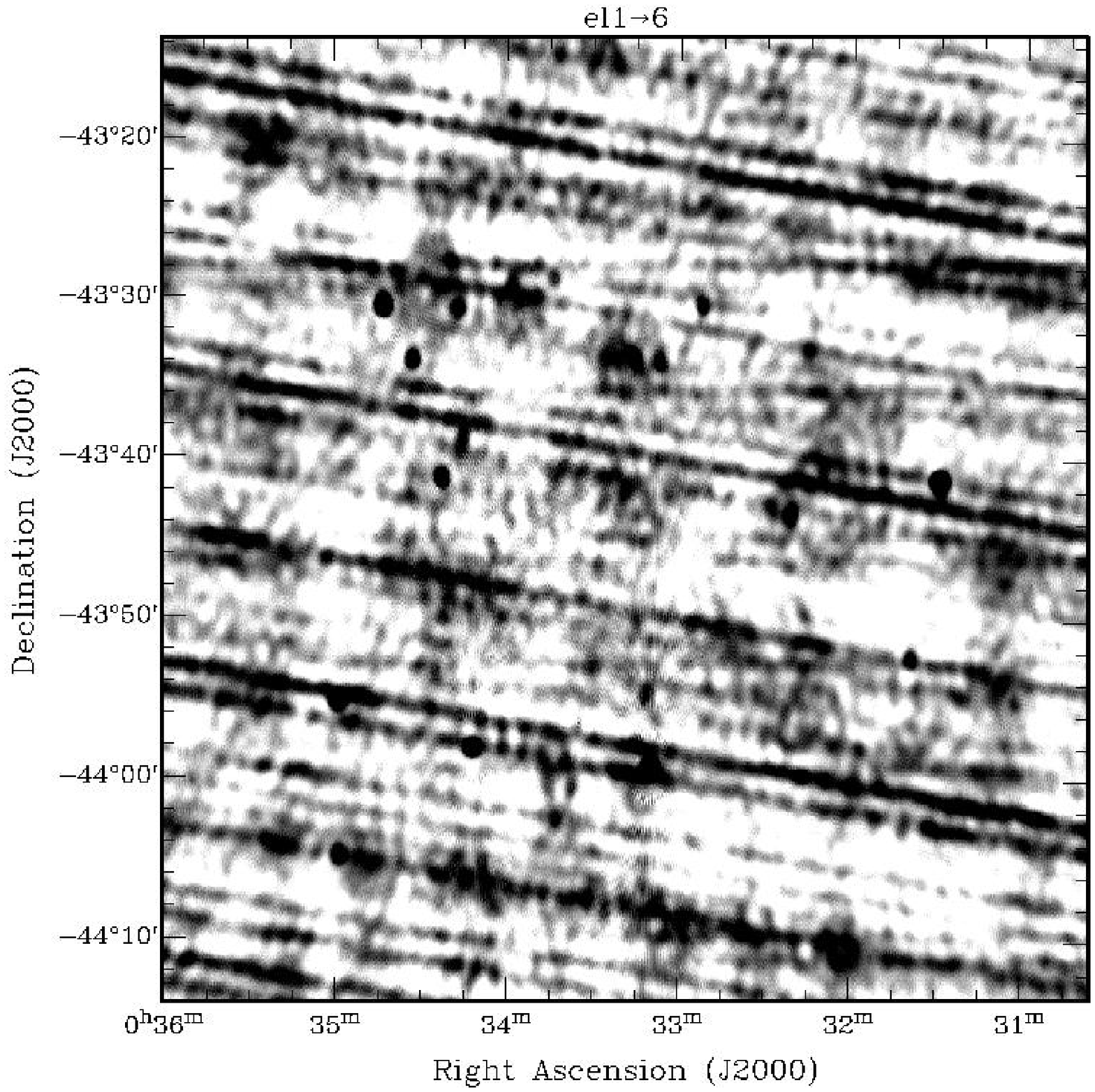}
\vspace{0.5cm}
\includegraphics[width=\linewidth]{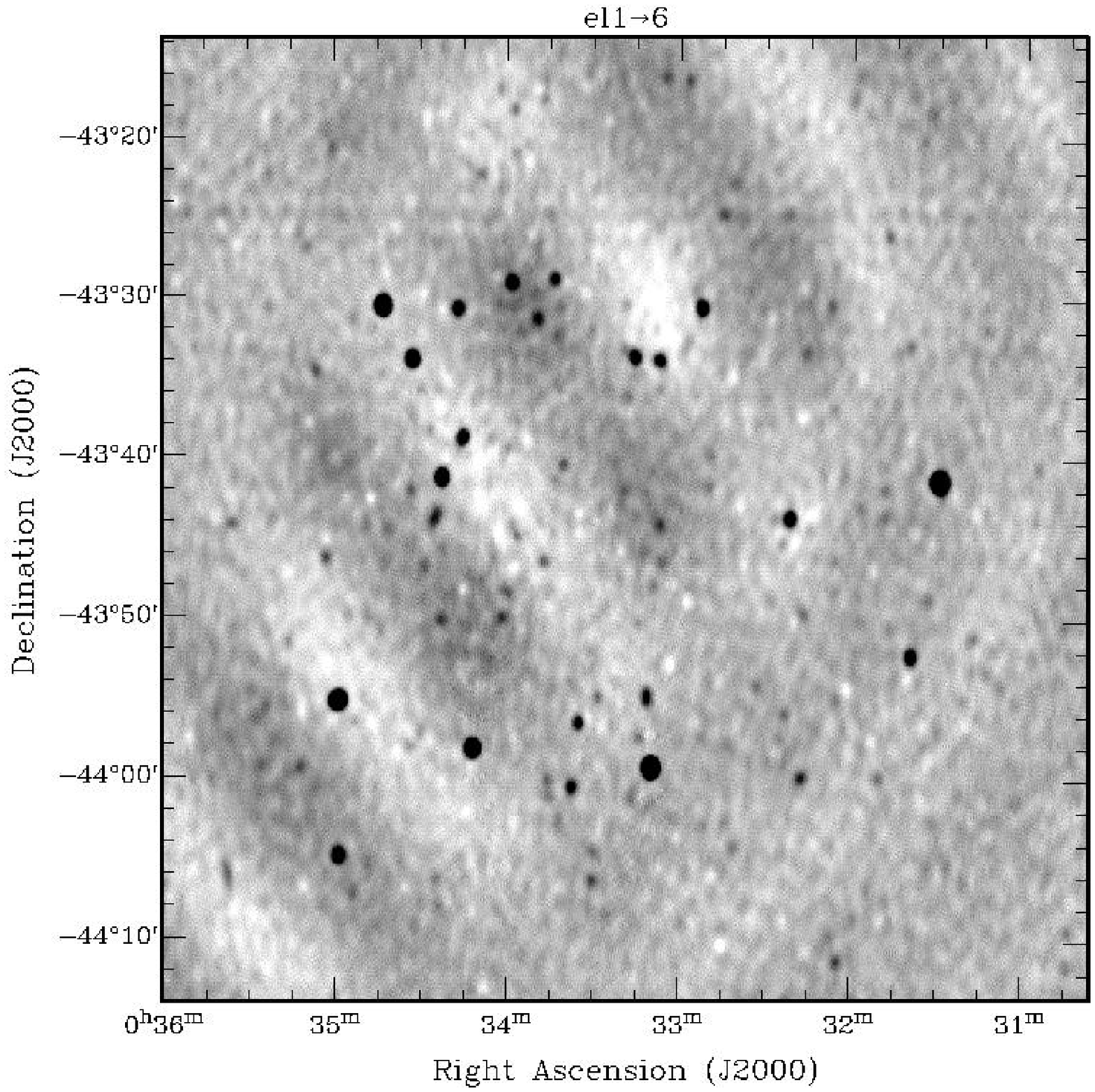}
\caption{Images made from a 1.4\,GHz ATCA detection
experiment with a net integration time of 25\,min. The effective
bandwidth is 208\,MHz, and the data have been calibrated in Miriad
according to the Miriad User Guide, then phase self-calibrated, and
imaged. The same colour transfer function has been used for the two
images. {\it Top:} No flagging was carried out, and the noise in the
image is 0.5\,mJy, approximately four times the thermal noise limit of
0.13\,mJy (as reported by the Miriad task ``invert''). {\it Bottom:}
The same data, but 10.6\,\% of the data have been flagged using
Pieflag. The image sensitivity is now 0.16\,mJy, and many more sources
than without flagging are clearly visible.}
\label{fig:images}
\end{center}
\end{figure}

%\end{multicols}

%\begin{thebibliography}{}
%% References are listed as in the following example, for more examples, please
%% see the PASA Style Guide
%\bibitem[Smith, Jones, \& Brown(Year)Smith et al.]{example}Smith, A.~B., Jones,
%C.~D., Brown, E.~F. Year, Journal, Volume, Page
%\end{thebibliography}

\bibliography{refs}

\begin{thebibliography}{2}
\expandafter\ifx\csname natexlab\endcsname\relax\def\natexlab#1{#1}\fi

\bibitem[{{Ellingson}(2004)}]{Ellingson2004}
{Ellingson}, S.~W. 2004, Experimental Astronomy, 17, 261

\bibitem[{{Sault} {et~al.}(1995){Sault}, {Teuben}, \& {Wright}}]{Sault1995}
{Sault}, R.~J., {Teuben}, P.~J., \& {Wright}, M.~C.~H. 1995, in ASP Conf. Ser.
  77: Astronomical Data Analysis Software and Systems IV, 433--+

\end{thebibliography}

\end{document}